\magnification=\magstep1
\centerline{\bf Of Course Muons Can Oscillate}
\medskip
\centerline{Y. N. Srivastava and A. Widom}
\centerline{Physics Department, Northeastern University, Boston MA 02115}
\centerline{and}
\centerline{Physics Department \& INFN, University of Perugia, Perugia, Italy}
\vskip .5in
\centerline{\it ABSTRACT}
\medskip
Recent theoretical claims not withstanding, muons can and do oscillate. 
Muons produced in association with neutrinos (if masses and mixing angles 
are nonzero) exhibit a joint oscillating spatial distribution.
The possible use of muon oscillations  as a probe of neutrino 
mass and mixing parameters is discussed using very simple physical 
arguments. Space-time oscillations in the secondary decay vertices of  
muons (produced by pion decay) persist after summing over all 
undetected neutrinos. 

\vskip .5in
\centerline{\bf 1. Introduction}
\medskip
  
Classic precision measurements of $(g-2)$ have been made for electrons[1] 
and muons[2-5]. For the latter case, muons are injected into a ring with 
a uniform applied magnetic field ${\bf B}$. The experimental quantum 
survival probability of the muon in the ring to decay, (via 
$\mu^- \to e^- +\bar{\nu}_e+\nu_\mu $) with a detected 
electron energy above a threshold value, has been fit to the 
theoretical oscillating form 
$$
P_{\mu^- \to e^-+\nu_\mu+\bar{\nu}_e}(t)=e^{-Mc^2 \Gamma t/E}
\Big({1+A \cos(\Omega t+\phi)\over 1+Acos\phi}\Big). \eqno(1.1)
$$ 
In Eq.(1.1), $M$, $E$, and $\Gamma^{-1}$ represent (respectively) the 
mass, energy, and intrinsic lifetime of the muon; $t$ is time in the 
laboratory reference frame, and 
$
\Omega =\big((g-2)eB/(2Mc)\big)  
$ 
is the experimental frequency measured in the muon survival probability 
due to quantum mechanical amplitude interference. Eq.(1.1) has been 
previously derived by us from the Dirac equation[6].

In view of the fact that the theoretical muon decay oscillations in 
Eq.(1.1) have been verified by some of the most precise experiments ever 
performed in high energy physics, it is sad to read a recent statement 
by Dolgov, Morozov, Okun and Schepkin[7] that ``Muons do not oscillate.''
The seminal muon oscillation work by Russian chemists[8] at their 
own institute in Moscow was ignored. Some incorrect 
physics in the work of Dolgov, Morozov, Okun and Schepkin[7], 
including (but not limited to) the fact that their equations are not  
dimensionally correct, is summarized in the Appendix.

Our purpose is to formulate the problem of neutrino induced muon 
oscillations using the notion of ``quantum beats'' which are discussed 
in recent text books[9-10] on quantum mechanics. The basic notion of a 
quantum mechanical decay exhibiting quantum beat oscillations may be 
formulated as follows: (i) The survival amplitude at time $t$ for a 
normalized state $\big|\Psi \big>$ is given by 
$$
{\cal S}(t)=\big<\Psi \big|e^{-iHt/\hbar}\big|\Psi \big>. \eqno(1.2)
$$   
(ii) The probability density for the state to have energy $E$ is 
given by 
$$
\rho (E)=\big<\Psi \big|\delta (E-H)\big|\Psi \big>. \eqno(1.3)
$$ 
From Eqs.(1.2) and (1.3) it follows that the survival amplitude for 
a state is the Fourier transform of the energy probability density 
$$
{\cal S}(t)=\int \rho (E) e^{-iEt/\hbar} dE, \eqno(1.4) 
$$
yielding the survival probability 
$$
P(t)=\big|{\cal S}(t)\big|^2=
\int \int \rho (E) \rho (E^\prime ) 
e^{-i(E-E^\prime )t/\hbar} dE dE^\prime . \eqno(1.5)  
$$  
Eqs.(1.2)-(1.5) are exact. {\it If the energy probability density 
$\rho (E) $ may be written as a sum of peaked resonances, then it 
follows rigorously that the survival probability $P(t)$ exhibits 
quantum beat oscillations in time.} The proof follows directly from 
Eq.(1.5).  

Eq.(1.5) (in a fixed reference frame) and its more general 
Lorentz invariant forms of expression are the basis of the discussion 
which follows for the theory of neutrino mass mixing induced 
muon decay oscillations. In Sec.2 quantum beat oscillations are 
discussed using the standard formalism successfully applied 
to atomic and molecular experimental decays. In Sec.3, the Lorentz 
invariant version of the theory is developed. In Sec.4 the four 
momentum distribution for the muon in the reaction  
$\pi^+\to \mu^++ \nu_\mu $ is discussed. In a model for which the 
neutrino has three possible mass eigenstates, and in which there is 
mixing, the (incoherent) momentum distributions obtained by us 
are identical to those conventionally used by others to analyse previous 
experiments. In Sec.5, the surival probability of the muon is 
computed and the interference effects (which are present if the 
neutrinos have both mass and flavor mixing) are 
discussed. In the concluding Sec.6, the rigorous relationship 
between the four momentum probability distribution and the survival 
probability is reviewed. An incoherent sum of the four 
momentum probability distribution may nevertheless exhibit quantum beat 
oscillations in the survival probability.    

\medskip 
\centerline{\bf 2. Quantum Beat Oscillations in Decays}
\medskip  

Consider the case of quantum beats in a decay. Suppose that the initial 
state of the system under consideration has the form 
$$
\big|\Psi \big>=c_1\big|\Psi_1 \big>+c_2\big|\Psi_2 \big>, \eqno(2.1)
$$  
with the assumption that $\big|\Psi_1 \big>$ and $\big|\Psi_2 \big>$ have 
no common decay modes. Explicitly, 
$$
\big<\Psi_2\big|e^{-iHt/\hbar}\big|\Psi_1\big>=
\sum_n \big<\Psi_2\big|n,out \big>
e^{-iE_{n,out}t/\hbar}\big<n,out\big|\Psi_1\big>=0, \eqno(2.2) 
$$
so that  
$$
\big<\Psi_2\big|\delta (E-H)\big|\Psi_1\big>=0,\ \ 
\big<\Psi_1\big|\delta (E-H)\big|\Psi_2\big>=0. \eqno(2.3)
$$
Eqs.(1.3), (2.1) and (2.3) imply
$$
\rho (E)=\big|c_1\big|^2\rho_1 (E)+\big|c_2\big|^2\rho_2 (E), \eqno(2.4)
$$
where
$$
\rho_1(E)=\big<\Psi_1 \big|\delta (E-H)\big|\Psi_1 \big>,\ \ 
\rho_2(E)=\big<\Psi_2 \big|\delta (E-H)\big|\Psi_2 \big>. \eqno(2.5) 
$$

Since the two states $\big|\Psi_1 \big>$ and $\big|\Psi_2 \big>$ have no 
common decay modes, the energy probability distribution $\rho (E)$ for 
the state $\big|\Psi \big>$ is a {\it  superposition of probabilities}  
of the energy probability densities 
$\rho_1(E)$ of state $\big|\Psi_1 \big>$ and     
$\rho_2(E)$ of state $\big|\Psi_2 \big>$, including the probabilities 
$\big|c_j\big|^2=\big|\big<\Psi_j \big|\Psi \big>\big|^2$. 

The superposition of energy probability densities in Eq.(2.4) by no 
means implies that there is a superposition of survival probabilities 
in time. In fact, Eqs.(1.4) and (2.4) imply a superposition of survival 
amplitudes having the form 
$$
{\cal S}(t)=\big|c_1\big|^2{\cal S}_1(t)
+\big|c_2\big|^2{\cal S}_2(t). \eqno(2.6)
$$
If the two states decay into different channels according to an 
exponential decay law 
$$
{\cal S}_1(t)=e^{-\Gamma_1 t/2}e^{-iE_1t/\hbar},\ \ 
{\cal S}_2(t)=e^{-\Gamma_2 t/2}e^{-iE_2t/\hbar}, \eqno(2.7) 
$$
then the survival probability 
$$
P(t)=|{\cal S}(t)|^2=\big|c_1\big|^4 e^{-\Gamma_1 t}+
\big|c_2\big|^4 e^{-\Gamma_2 t}+2\big|c_1\big|^2\big|c_2\big|^2 
e^{-(\Gamma_1+\Gamma_2)t/2}cos(\omega_{12} t), \eqno(2.8)
$$
where 
$$
\hbar \omega_{12} =E_1-E_2 . \eqno(2.9)
$$

Eqs.(2.8) and (2.9) are fundamental equations for the quantum 
beat oscillations in decays which constitute (for example)  
commonplace experimental phenomena in atomic and molecular physics 
and chemistry.  
\medskip 
\centerline{\bf 3. Lorentz Invariant Space-Time Oscillations }
\medskip  

The amplitude that a normalized state $\big|\Psi \big>$ 
describing a physical system in the neighborhood of the space-time 
origin survives a displacement to the neighborhood of the space-time 
point $x$ may be defined by 
$$
{\cal S}(x)=\big<\Psi \big|e^{iP\cdot x/\hbar }\big|\Psi \big>, 
\eqno(3.1)
$$
where 
$$
P=({\bf P},H/c) \eqno(3.2)
$$
is the operator four momentum. Note that Eq.(3.1) reduces to 
Eq.(1.2) for decays in the center of mass frame[11] for which  
the three momentum may be set to zero.

In a Lorentz covariant manner, one may start from Eq.(3.1) and 
write 
$$
{\cal S}(x)=\int \sigma (k)e^{ik\cdot x}d^4 k, \eqno(3.3)
$$
where the four momentum distribution is given by  
$$
\sigma (k)=\Big<\Psi \Big|\delta^{(4)}\Big(k-{P\over \hbar}\Big)
\Big|\Psi \Big>. \eqno(3.4)
$$

If the initial state is in a superposition of finitely many states, 
$$
\big|\Psi \big>=\sum_j c_j \big|\Psi_j \big>, \eqno(3.5)
$$
no two of which have common decay modes, then the four momenta 
probability densities are given by the (incoherent) superposition 
weighted by the probabilities $|c_j|^2$  
$$
\sigma (k)=\sum_j \big|c_j\big|^2 \sigma_j(k), \eqno(3.6)
$$
where 
$$
\sigma_j (k)=\Big<\Psi_j \Big|\delta^{(4)}\Big(k-{P\over \hbar}\Big)
\Big|\Psi_j \Big>. \eqno(3.7)
$$
Eqs.(3.6) and (3.7) are the Lorentz invariant form of Eqs.(2.4) 
and (2.5).
The survival amplitude follows from Eqs.(3.3) and (3.6) to be 
of the superposition of amplitudes form 
$$
{\cal S}(x)=\sum_j \big|c_j\big|^2 {\cal S}_j(x). \eqno(3.8) 
$$
Clearly, the survival probability exhibits quantum interference 
effects 
$$
P(x)=\big|{\cal S}(x)\big|^2=
\sum_{i,j}\big|c_i\big|^2\big|c_j\big|^2 
{\cal S}^*_i(x){\cal S}_j(x), \eqno(3.9)
$$
and quantum beat oscillations in decays have been described in 
a Lorentz invariant form.  
\medskip
\centerline{\bf 4. Muon Momentum Distributions 
for $\pi^+\to \mu^++ \nu_\mu $}
\medskip

We consider a massive neutrino model with mixing and define the 
following parameters: For three flavors of neutrino 
($\nu_e$, $\nu_\mu $, and $\nu_\tau $) there is a 
rotation matrix 
$$
\pmatrix{\nu_e \cr \nu_\mu \cr \nu_\tau }=
\pmatrix 
{ R_{e1} & R_{e2} & R_{e3} \cr
R_{\mu 1} & R_{\mu 2} & R_{\mu 3} \cr
R_{\tau 1} & R_{\tau 2} & R_{\tau 3} }
\pmatrix{n_1 \cr n_2 \cr n_3 }, \eqno(4.1)
$$
where $n_1$, $n_2$, and $n_3$ denote neutrino mass states with masses 
$m_1$, $m_2$, and $m_3$ respectively. In the reaction 
$$
\pi^+\to \mu^+ + \nu_\mu , \eqno(4.2)
$$
the four momentum distribution for the $\mu^+ $ has the form of 
Eq.(3.6); i.e. 
$$
\sigma_{muon}(k)=\sum_{j=1}^3 \big|R_{\mu j} \big|^2 
\sigma_{muon}(k,m_j), \eqno(4.3)
$$
where $\sigma_{muon}(k,m_j)$ is the normalized four momentum 
distribution of the $\mu^+ $ obtained {\it if} the original muon 
produced in the pion decay were to recoil against a neutrino of 
mass $m_j$. 

Note that the four momentum probability density for the muon in 
Eq.(4.3), is an incoherent probability superposition. We have already 
summed over the unobserved neutrino final states, 
both flavor and momentum. The incoherent four momentum distribution for 
the muon has long been understood (and employed) in the analysis of 
experimental pion decays as a probe of possible neutrino masses[12]. 
Unfortunately, the previous experimental resolution for measured 
muon momentum distributions did {\it not} allow for the definitive 
detection of any finite neutrino masses. However, the resolution 
did allow for an improved determination[12] of the pion mass. 
\medskip 
\centerline{\bf 5. Muon Survival Probability 
from $\pi^+\to \mu^++ \nu_\mu $}
\medskip

From Eqs.(3.6), (3.8) and (4.3), the muon survival amplitude against  
$\mu^+\to e^+ +\bar{\nu}_\mu +\nu_e$ from muons produced in a pion 
decay is given by 
$$
S_{muon}(x)=\sum_{j=1}^3 \big|R_{\mu j} \big|^2 S_{muon}(x,m_j), 
\eqno(5.1)
$$ 
where 
$$
S_{muon}(x,m_j)=\int \sigma_{muon} (k,m_j)e^{ik\cdot x}d^4 k  
\eqno(5.2)
$$
is the survival amplitude of the $\mu^+ $ obtained {\it if} the muon 
were to recoil against a neutrino of mass $m_j$. The survival amplitude 
$S_{muon}(x,m_j)$ is very well known; i.e.  
$$
S_{muon}(x,m_j)\approx 
e^{ip_j\cdot x/\hbar }e^{-\Gamma \sqrt{-x^2}/2c}. \eqno(5.3)
$$

In the pion rest frame $p_j=({\bf p}_j,\sqrt{M_\mu^2 c^2+|{\bf p}_j|^2})$, 
$$
{|{\bf p}_j|\over c}=
{\sqrt{\big(M_\pi^2-(M_\mu +m_j)^2 \big)\big(M_\pi^2-(M_\mu -m_j)^2 \big)}
\over 2M_\pi}. \eqno(5.4)
$$
In any reference frame, the survival probability for the muon follows from 
Eqs.(5.1) and (5.3) to be 
$$
P_{muon}(x)=\big|S_{muon}(x)\big|^2\approx 
e^{-\Gamma \sqrt{-x^2}/c} \sum_{j=1}^3\sum_{l=1}^3 
\big|R_{\mu j}\big|^2 \big|R_{\mu l}\big|^2  
cos\big((p_j-p_l)\cdot x/\hbar \big). \eqno(5.5)
$$
The presence of quantum beat oscillations in the muon survival 
probability, see Eqs.(5.4) and (5.5), is the central result of this work. 
The incoherent sum in the four 
momentum probability distribution in Eq.(4.3) exhibits 
quantum beat oscillations in the survival probability in Eq.(5.5).

\medskip 
\centerline{\bf 6. Conclusion}
\medskip

In the pion decay $\pi^+\to \mu^++ \nu_\mu$, the muon later also decays 
via $\mu^+\to e^+ +\bar{\nu}_\mu +\nu_e$. Assume that the neutrinos have 
both mass and flavor mixing. The four-momentum distribution 
$\sigma_{muon}(k)$ of the muon produced by the pion will be a 
probability superposition as in Eq.(4.3). This result is well known[12],  
and has already been employed in the analysis of pion decay experiments 
searching for neutrino masses. For the muon survival amplitude, the  
well known Eq.(4.3) yields Eq.(5.1) and for the survival probability 
Eq.(5.5). 

The interference effects are manifest in the survival probability 
Eq.(5.5). The secondary vertex from 
$\mu^+\to e^+ +\bar{\nu}_\mu +\nu_e$, distributed via 
$P_{muon}(x)$, may provide a more sensitive probe of neutrino masses 
and mixing than does the four momentum distribution via 
$\sigma_{muon}(k)$. Further details about oscillations may be found 
in our earlier works[13-16].

\vskip .5in
\centerline{\it APPENDIX}
\medskip 

In his lectures on quantum field theory, T. D. Lee[17] states 
``An equation in physics, say $A=B$, must satisfy the requirement that 
the dimension of $A$ be the same as that of $B$: 
$\big[A\big]=\big[B \big]$. This seemingly elementary requirement 
can serve the useful purpose of verifying the correctness of one's 
equations, especially after a long calculation.'' 

Almost all the central probability equations in the work of 
Dolgov, Morozov, Okun and Schepkin[7], hereafter abbreviated 
DMOS[7], are dimensionally inconsistent and thereby 
fail to satisfy T. D. Lee's elementary requirement for correct 
physical equations.

In Eq.(9), DMOS[7] define a dimensionless joint muon neutrino 
probability 
$$
P_{\nu_a}^A(x_\mu,x_\nu)=\beta_{1a}^2 + \beta_{2a}^2 
+2 \beta_{1a}^2\beta_{2a}^2 cos(\phi_1-\phi_2) \eqno(DMOS 1)
$$
where $\beta_{ja}$ are seen to be dimensionless in Eq.(11) 
of DMOS[7]. In Eq.(25), DMOS[7] define a single muon 
probability 
$$
P_\mu ^B(x_\mu )=\sum_a\int P_{\nu_a}^A(x_\mu,x_\nu) d{\bf x}_\nu, 
\eqno(DMOS 2). 
$$
The quantity on the left hand side of Eq.(DMOS2) is alleged to be 
dimensionless (in fact unity)  from Eq.(27) of DMOS. The right hand 
side of Eq.(DMOS2) has dimensions of $\big[length^3\big]$. By 
virtue of dimensional analysis alone, their result for the muon 
probability distribution is wrong. The integrations are simply incorrect. 
Errors evident from dimensional considerations are also present 
(for example) in Eq.(26) of DMOS[7]. 

The dimensional errors are symptomatic of a confusion 
in the starting wave function Eq.(6) of DMOS[7] 
$$
\psi_{p_\pi}(x_\mu, x_\nu |x_i)=|\mu >
(e^{-i\phi_1}cos\theta|\nu_1>+e^{-i\phi_2}sin\theta|\nu_2>). 
\eqno(DMOS3)
$$  
Imagine a world in which one of the neutrinos has a large mass and the 
other neutrino has virtually zero mass. An energetic pion decays at CERN 
so that the muon ``ket'' which recoils from the heavy neutrino decays in 
Switzerland and the muon ``ket'' which recoils from the massless neutrino 
decays in France. In the DMOS[7] work the ``ket''  in France 
and the ``ket'' in Switzerland are {\it identical}, i.e. $|\mu >$ in 
Eq.(DMOS3). We prefer to think that 
$<\mu_{France}|\mu_{Switzerland}>=0$. 
At any rate, without this physical notion there is more than a little 
difficulty in properly normalizing wave functions.   

\vskip .5in
\centerline{\bf References}
\medskip
\par \noindent
1. P. Kusch and H. M. Foley, {\it Phys. Rev.} {\bf 72}, 250 (1948). 
\par \noindent
2. G. Charpak, et. al. {\it Il Nuovo Cimento} {\bf 37}, 1241 (1965).
\par \noindent
3. J. Bailey, et. al. {\it Il Nuovo Cimento} {\bf 9A}, 369 (1972).
\par \noindent
4. F. Combley and E. Picasso, {\it Phys. Rep.} {\bf 14}, 1 (1974).
\par \noindent
5. J. Bailey, et. al. {\it Nucl. Phys.} {\bf B150}, 1 (1979).
\par \noindent
6. A. Widom ansd Y. N. Srivastava, ``Charged Lepton Oscillations and 
(g-2) measurements'', hep-ph/9612290 13 Dec 1996.
\par \noindent
7. A. D. Dolgov, A. Yu.Morozov, L. B. Okun and M.G. Schepkin ``Do Muons 
Oscillate?'', hep-ph/9703241 5 Mar 1997. 
\par \noindent
8. V. I. Goldanski and V. G. Firsov, {\it Ann. Rev. Phys. Chem } {\bf 22}, 
209 (1971).
\par \noindent 
9. G. Greenstein and A. G. Zajonc, ``The Quantum Challenge'', pp. 92-5, 
Jones and Bartlett Publishers, Boston (1997)
\par \noindent
10. M. P. Silverman, ``More Than One Mystery: Explorations in Quantum 
Interference'', Chapt. 4, Springer-Verlag Berlin (1995). 
\par \noindent
11. M. L. Goldberger and K. M. Watson, ``Collision Theory'', 
Chapt. 8, John Wiley and Sons, New York (1964).
\par \noindent
12. R. Abela, et. al. {\it Phys. Lett.} {\bf B146}, 431 (1984).
\par \noindent
13. Y. N. Srivastava, A. Widom and E. Sassaroli, 
{\it Phys. Lett.} {\bf B344}, 436 (1995).
\par \noindent
14. Y. N. Srivastava, A. Widom and E. Sassaroli,
{\it ``Real and Virtual Strange Processes''} hep-ph/9507330, 
Proceedings of Workshop on Physics and Detectors for DA$\Phi$NE 
Laboratori Nazionali di Frascati, Frascati (1995).  
\par \noindent
15. Y. N. Srivastava, A. Widom and E. Sassaroli,
{\it ``Lepton Oscillations''} hep-ph/9509261.
\par\noindent
16. Y. N. Srivastava, A. Widom and E. Sassaroli, 
{\it ``Associated Lepton Oscillations''}, Proceeding of the Conference 
``Results and Perspectives in Particle Physics'' Editor M. Greco, 
Les Rencontres de Physique de la Vallee d' Aoste, La Thuile (1996).
\par \noindent
17. T. D. Lee ``Particle Physics and Introduction to Field 
Theory'', page 2, Harwood Academic Publishers, London (1981).

\bye